\documentclass[aps,pra,superscriptaddress,notitlepage]{revtex4-1}
\usepackage{amsmath,amssymb}
\usepackage{graphicx}
\usepackage[dvipsnames]{xcolor}
\usepackage{comment}
\usepackage{upgreek}
\usepackage{natbib}
\usepackage{cases}

\renewcommand{\bm}[1]{\boldsymbol{\mathbf{#1}}}

\providecommand*{\pp}[3][]{\frac{\partial^{#1}#2}{\partial #3^{#1}}}
\providecommand*{\dd}[3][]{\frac{\mathrm{d}^{#1}#2}{\mathrm{d} #3^{#1}}}
\providecommand*{\DD}[3][]{\frac{\mathrm{D}^{#1}#2}{\mathrm{D} #3^{#1}}}

\providecommand*{\rmd}{\mathrm{d}}

\providecommand*{\m}{\displaystyle}
\providecommand*{\upi}{\uppi}

\newcommand{\bv}{\bm{v}}
\newcommand{\bvhat}{\hat{\bv}}
\newcommand{\bsigma}{\bm{\sigma}}
\newcommand{\bsigmahat}{\hat{\bsigma}}
\newcommand{\ex}{\bm{e}_x}
\newcommand{\ey}{\bm{e}_y}
\newcommand{\ez}{\bm{e}_z}
\newcommand{\intinf}{\int_{-\infty}^{\infty}}
\newcommand{\hatsig}{\hat{\sigma}}

\AtBeginDocument{%
  \renewcommand*\O{\mathcal{O}}  % Big O
}

\begin{document}
\preprint{}

\title{Rotation of an immersed cylinder sliding near a thin elastic coating}


\author{Bhargav Rallabandi}
\email{vbr@princeton.edu}
\affiliation{Department of Mechanical and Aerospace Engineering, Princeton University, Princeton,
	New Jersey 08544, USA}
\author{Baudouin Saintyves}
\affiliation{John Paulson School of Engineering and Applied Sciences, Harvard University, Cambridge, MA 02138, USA}
\author{Theo Jules}
\affiliation{John Paulson School of Engineering and Applied Sciences, Harvard University, Cambridge, MA 02138, USA}
\affiliation{D\'{e}partment de Physique, \'{E}cole Normale Sup\'{e}rieure, PSL Research University, 75005 Paris, France}
\author{Thomas Salez}
\affiliation{Department of Mechanical and Aerospace Engineering, Princeton University, Princeton, New Jersey 08544, USA}
\affiliation{Laboratoire de Physico-Chimie Th\'{e}orique, UMR CNRS Gulliver 7083,
	ESPCI Paris, PSL Research University, 75005 Paris, France}
\affiliation{Global Station for Soft Matter, Global Institution for Collaborative Research and Education,
Hokkaido University, Sapporo, Hokkaido 060-0808, Japan}
\author{Clarissa Sch\"{o}necker}
\affiliation{Max Planck Institute for Polymer Research, 55128 Mainz, Germany}
\author{L. Mahadevan}
\affiliation{John Paulson School of Engineering and Applied Sciences, Harvard University, Cambridge, MA 02138, USA}
\author{Howard A. Stone}
\email{hastone@princeton.edu}
\affiliation{Department of Mechanical and Aerospace Engineering, Princeton University, Princeton,
	New Jersey 08544, USA}
%\thanks{}


\date{\today}

\begin{abstract}
	{\color{black}  It is known that an object translating parallel to a soft wall in a viscous fluid produces hydrodynamic stresses that deform the wall, which, in turn, results in a lift force on the object. Recent experiments with cylinders sliding under gravity near a soft incline, which confirmed theoretical arguments for the lift force, also reported an unexplained steady-state rotation of the cylinders [Saintyves et al. \emph{PNAS} 113(21), 2016]. Motivated by these observations, we show, in the lubrication limit, that an infinite cylinder that translates in a viscous fluid parallel to a soft wall at constant speed and separation distance must also rotate in order to remain free of torque. Using the Lorentz reciprocal theorem, we show analytically that for small deformations of the elastic layer, the angular velocity of the cylinder scales with the cube of the sliding speed. These predictions are confirmed numerically. We then apply the theory to the gravity-driven motion of a cylinder near a soft incline and find qualitative agreement with the experimental observations, namely that a softer elastic layer results in a greater angular speed of the cylinder. }
\end{abstract}

%\keywords{}

\maketitle

\section{Introduction}

{\color{black} The topic of hydrodynamically mediated interactions between elastic objects is widely researched and has applications in engineering, biophysics, and geophysics. }Studies have included applications to roll coating \citep{coy88_rollcoating} and printing \citep{yin05_cavityflexiblewall}, collisions between suspended particles \citep{dav86}, the rheology of polymer-bearing surfaces \citep{sek93}, the behavior of vesicles near walls \citep{abk02} and red blood cells in capillaries \citep{fre14_rev_blood}, the mobility of suspended objects near elastic membranes \citep{dad17_axisym,dad16}, lubrication in cartilagenous joints \citep{gro78_cartilage,hou92_cartilage} and the mechanics of seismic faults \citep{biz12_faults}. Often, the salient coupling between elasticity and flow is most effective when surfaces are in near contact, allowing the well-developed literature on lubrication flows \citep{gol67a,jef81} to be suitably adapted to include the boundary deformation. Recent studies in the context of lubrication flows with deformable boundaries have quantified the roles of fluid compressibility \citep{bal10_compressible}, inertia of the fluid and the elastic medium \citep{cla11}, and the effect of a viscoelasticity of the surfaces \citep{pan15}.

A well-known consequence of the reversibility of Stokes flows is that a rigid symmetric object such as a sphere or a cylinder that translates in a viscous fluid parallel to a rigid wall at zero Reynolds number experiences no hydrodynamic force normal to the wall. However, if either surface is soft, the parallel translation of such an object induces stresses in the fluid that deform the surfaces. The surface deformation breaks symmetry and drives a secondary Stokes flow that can produce a lift force on the object \citep{sek93,sko04}. %Thus, a rigid sphere or cylinder, when acted upon by an external force parallel to a soft wall, must translate with a non-zero component of its velocity normal to the wall in order to remain free of net force.
The consequences of such a lift force have been observed experimentally and described theoretically in a number of different configurations \citep{sek93,abk02,sai16}, including those in which transient dynamics are important \citep{wee06_transient,bal10_compressible,sal15}.

By and large, the focus has been on effects of boundary elasticity on the translation of a rigid object suspended in a viscous fluid \citep{dav86,sko05,sno13,pan15}. Recent experiments \citep{sai16} measuring the influence of elasticity on the motion of a rigid cylinder sliding parallel to a soft wall, however, also noted a spontaneous rotation of the cylinder. On the other hand, asymptotic theory that is accurate to leading order in the deformation of the elastic wall does not predict an elasticity-dependent rotation of a rigid cylinder that translates parallel to a soft wall at steady state, even though transient rotation is indeed predicted \citep{sal15}. 

{\color{black}
In this article, we show that the coupling between fluid flow and elastic deformation induces a self-sustained rotation of a nearby rigid object by developing a higher-order asymptotic analysis that accounts for both the force and the torque balances. Focusing on the case of an infinite cylinder sliding parallel to a soft elastic layer, we use the Lorentz reciprocal theorem along with previously known results \citep{sek93,sko04,sko05} to show formally, in the limit of small elastic deformations, that a torque-free translating cylinder rotates at quadratic order in the deformation amplitude of the layer. These results are in agreement with numerical solutions with finite deformation amplitudes of the layer. %Consequently, we infer that an infinite cylinder that slides parallel to a soft wall without rotating can only do so under an externally applied torque that balances the elastohydrodynamic torque.
After first deriving a general expression for the cylinder's angular velocity in terms of its sliding speed, its radius and the dimensionless compliance of the elastic layer, we apply our result to the rotation of a cylinder free-falling under gravity near a soft incline.  {\color{black} The results of the analysis are in qualitative agreement with the experimental observations of \citet{sai16} for gravity-driven sliding near an incline: the angular velocity increases with the softness of the elastic layer.} 
We then discuss symmetry arguments that are generally applicable to systems with deformable boundaries. }

%{\color{black} Reword?}We compare the theoretical predictions with direct experimental measurements, and then develop symmetry arguments that are generally applicable to systems with deformable boundaries.

\section{Problem description and scaling}
\begin{figure}[t!]
	\centering
	\includegraphics[scale=1]{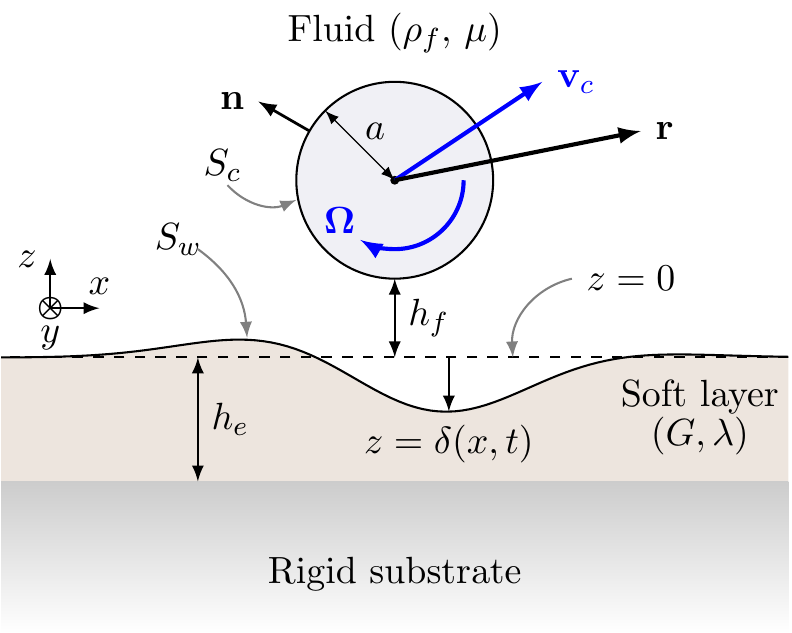}
	\caption{Sketch of the geometry and coordinate system, showing the translating and rotating cylinder and the outline of the deformed wall.}
	\label{FigSetup}
\end{figure}

We consider the motion of an infinite cylinder of radius $a$ in an incompressible viscous fluid of density $\rho_f$ and viscosity $\mu$. The cylinder translates and rotates near a wall, as shown in figure \ref{FigSetup}, under the action of an external force. The wall comprises a rigid substrate coated with a soft elastic layer of thickness $h_e$, with shear and bulk moduli $G$ and $\lambda$, respectively. We assume that the motion of the cylinder occurs in such a way that its axis remains tangent to the undeformed reference surface of the soft layer at all times. The smallest separation distance between the surface of the cylinder and the nominally flat (undeformed) reference surface of the soft layer is denoted by $h_{f}$.

We restrict our attention to two-dimensional motion in the $xz$ plane, so that the cylinder's translational velocity, when expressed in the laboratory frame is $\bm{v}_c = u_{s} \ex + w_c \ez$, and its angular velocity is $\bm{\Omega} = \Omega\, \ey$ (see figure \ref{FigSetup}). The motion of the cylinder drives a fluid flow with a velocity $\bm{v}(\bm{x},t)$ and an associated stress field $\bm{\sigma}(\bm{x},t) = -p \bm{I} + \mu( \nabla \bm{v} + \nabla \bm{v}^T)$, where $p(\bm{x},t)$ is the pressure. If the Reynolds number is small, the flow is described by the continuity and Stokes equations
%\begin{subequations} \label{stokesflow}
\begin{align} \label{stokesflow}
&\nabla \cdot \bm{v} = 0 \quad  \mbox{and} \quad \nabla \cdot \bm{\sigma}  = \bm{0}\,.
\end{align}
%\end{subequations}
The stresses due to the flow deform the soft wall. We denote this deformation by $\delta(x,t)$, and use the convention that $\delta$ is negative under compression. The relaxation of stresses in the soft material is assumed to be rapid relative to the characteristic time scale of the flow, so that the wall can be modeled as a purely elastic medium. Then, in the absence of inertia, the elastic wall responds instantaneously to the stresses applied on it by the flow. We also assume that the inertia of the cylinder is negligible. 

It is convenient to work in a frame of reference that translates with the velocity $\bm{v}_c$ of the cylinder, and whose origin at a time $t$ is at the point on the undeformed position of the wall closest to the surface of the cylinder (see figure \ref{FigSetup}). In these coordinates, the deformed wall is located at $z = \delta$, and the axis of the cylinder at $(x=0,z = h_f(t) + a)$. Henceforth it will be understood that all velocities, including the fluid velocity $\bm{v}(\bm{x},t) = u(\bm{x},t) \ex + w (\bm{x},t) \ez$, are measured relative to the translating frame of reference. Although the eventual focus is on steady-state parallel translation ($w_c = 0$), we first develop the theory in more general terms below.

Noting that a perfectly rigid wall, as measured in the translating reference frame, translates with velocity $-\bm{v}_c = (-u_c,-w_c)$, the kinematic condition at the deformed wall $z=\delta(x,t)$ can be expressed as $w = -w_c + \DD{}{t}[\delta(x, t)]$, where $\DD{}{t}$ is the material time derivative. Using the no-slip condition at the wall $u = -u_c$ the boundary conditions for the flow are therefore
\begin{subequations}\label{eq:bc}
	\begin{align} 
	\bm{v}  &= \bm{\Omega} \times \bm{r} \quad \mbox{on} \quad S_c, \\ 
	\bm{v}  &= -\bm{v}_c - \left(  u_c \pp{\delta}{x} - \pp{\delta}{t}\right) \ez \quad \mbox{on} \quad S_w ,
	\end{align}
\end{subequations}
where $\bm{r}$ is a position vector relative to the center of the cylinder (see figure~\ref{FigSetup}), and $S_c$ and $S_w$ represent the surfaces of the cylinder and the deformed wall, respectively. The condition (\ref{eq:bc}a) is a statement of no slip on the cylinder in the translating reference frame. The term involving $\pp{\delta}{t}$ in (\ref{eq:bc}b) is important in the transient dynamics of the cylinder, but vanishes for steady state wall-parallel translation \citep{wee06_transient}.

To complete the description of the problem, it is necessary to specify the relation between the fluid stresses and the wall deformation.  Here, we model the elastic deformation as being proportional to the applied local normal stress (i.e. the Winkler approximation), which is applicable for $\lambda/G = \O(1)$ and small deformations $|\delta| \ll h_e \ll \ell$, where $\ell$ is the characteristic length scale of the stress distribution along the wall \citep{johnson1987contact}. Within this approximation, the elastic deformation $\delta$ can be written as
\begin{equation} \label{delta}
    \delta(x,t) =  \frac{h_e}{2G + \lambda}  \bm{\sigma} \bm{:}\, \ez\ez \big|_{S_w}.
\end{equation}

We henceforth restrict ourselves to the lubrication limit ($h_f \ll a$) in which $\ell \equiv \sqrt{2 a h_f} $ is the characteristic length scale of the contact region \citep{jef81,sto05b}. {\color{black} Note that this places the restriction $|\delta|^2/a \ll h_e^2/a \ll h_f$ in order for \eqref{delta} to remain applicable, i.e. $h_f$ cannot be too small.} The elastohydrodynamic lift force on a sliding cylinder arising from the deformation of the wall in this limit is known in a number of different situations \citep{sek93,sko05,sno13,pan15}. The effect of elasticity on the steady-state rotation of the cylinder has, however, not been quantified systematically, although it has been observed experimentally \citep{sai16}. Here, we focus on developing an analytic theory for the induced elastohydrodynamic rotation of a torque-free infinite cylinder sliding parallel to a soft wall in the limit of small deformations. These predictions are confirmed by numerical solutions for finite deformations.

\subsection{Scaling}

Before developing a detailed theory, we first present scaling arguments for the rotation rate. In the lubrication limit, the normal stress is dominated by the pressure of the flow in the lubrication layer, so that $|\bm{\sigma} \bm{:}\, \ez\ez| \approx p \propto \mu u_c \ell/h_f^2$. The order of magnitude of the deformation of the soft layer can be estimated using (\ref{stokesflow}) and (\ref{delta}) as $\delta \propto \mu u_c \ell h_e/(h_f^2 (2G + \lambda))$. The characteristic scale of the deformation relative to the thickness of the lubrication layer $h_f$ is thus quantified by the dimensionless parameter
\begin{equation} \label{LambdaDef}
	\Lambda \equiv \frac{\mu u_c \ell h_e}{h_f^3 (2G + \lambda)} = \frac{\sqrt{2} \mu u_c a^{1/2} h_e}{h_f^{5/2} (2G + \lambda)}.
\end{equation}
Here,  $\Lambda$ can be interpreted either as a dimensionless elastic compliance or a characteristic amplitude of elastic deformation relative to the width of the lubrication gap, i.e. $|\delta|/h_f = \O(\Lambda)$ by construction.

For small deformations $(\Lambda \ll 1)$, the lift force per unit length on a cylinder sliding parallel to a soft wall is approximately $\Lambda \mu u_c a/h_f$ \citep{sko04}. {\color{black} On the contrary, a translating cylinder experiences zero torque up to leading order in $\Lambda$; this was shown to be true for a rigid wall ($\Lambda = 0$) by \citet{jef81} and at $\O(\Lambda)$ by \citet{sal15} in the lubrication limit. This does not, however, preclude rotation due to higher order contributions: the rotation rate of a torque-free sliding cylinder can be estimated by interpreting the zero net torque on the cylinder as a superposition of equal and opposite contributions due to translation and rotation.} Since the lubrication gap in between the cylinder and the deformed soft wall has thickness $(h_f + \delta)$, the torque per unit length on a purely sliding cylinder scales nominally as $\mu u_c a \ell/(h_f + \delta)$. Using a Taylor series expansion in $\Lambda$ (small deformations; $\delta/h_f = \O(\Lambda)$) and invoking the result that a sliding cylinder experiences no torque  up to $\O(\Lambda)$, we conclude that the leading-order elastohydrodynamic torque (per unit length) due to sliding scales as $\mu u_c a \ell \Lambda^2/h_f$. On the other hand, the torque per unit length on a purely rotating cylinder for small deformations is $\O(\mu \Omega a^2\ell/h_f)$. Balancing the two contributions and using \eqref{LambdaDef} results in the scaling relation
\begin{equation} \label{scaling}
	\frac{a\Omega}{u_c} \propto \Lambda^2 \propto \frac{\mu^2 u_c^2 a h_e^2}{h_f^5 (2G + \lambda)^2} .
\end{equation} 
Therefore, the rotation of the cylinder is expected to occur at $\O(\Lambda^2)$, in contrast with the lift force, which depends linearly on $\Lambda$ at leading order \citep{sek93,sko04}. We quantify this rotational scaling formally in the sections below by a systematic application of the Lorentz reciprocal theorem for Stokes flow, which is in agreement with numerical solutions. We then apply our general result to the case of buoyancy-driven motion of a cylinder near an inclined wall coated with a soft layer.

\section{Theory}
\subsection{A reciprocal relation for a cylinder rotating near a deformable wall} \label{sec:Recip}
First, we develop a reciprocal relation for the angular velocity of a torque-free cylinder near a weakly deformable wall. While the focus of this article is on the elastic deformations that are described by the approximation \eqref{delta} and the lubrication limit $h_f \ll a$, the methodology described below is generally applicable to flow problems involving boundary deformations that are small relative to the smallest geometric length scale.

For small deformations $|\delta| \ll h_f$, it is appropriate to express the no-slip condition (\ref{eq:bc}b) on the deformed wall $S_w$ by a perturbation of the domain about the undeformed position of the wall, see e.g. \citet{hin91_book}. We note that since $\delta/h_f = \O(\Lambda)$, a formal perturbation expansion for small deformations involves powers of $\Lambda$. Due to the generality of the technique employed in this section, which only relies on $|\delta|/h_f \ll 1$ independent of the constitutive law relating stresses to the deformation of the wall, we will avoid an explicit expansion in powers of the dimensionless parameter $\Lambda$ at this stage, but instead directly expand in terms of dimensional quantities. Writing $\bm{v}(\bm{x})$ in (\ref{eq:bc}b) using a Taylor expansion about $z=0$, we obtain the equivalent condition 
\begin{equation} \label{eq:vbcz0}
	\bm{v}  = -\bm{v}_c - \left(u_c \pp{\delta}{x} -  \pp{\delta}{t}\right) \ez - \delta \pp{\bm{v}}{z} - \frac{\delta^2}{2} \pp[2]{\bm{v}}{z} + \O\left(\delta^3 \pp[3]{\bm{v}}{z}\right) \quad \mbox{on} \quad S_{w0},
\end{equation}
where $S_{w0}$ denotes the \emph{undeformed} wall position $z=0$. Note that the error term in the expansion above is $\O(\Lambda^3)$ relative to the leading-order translational velocity of the wall, $-\bm{v}_c$.

To utilize the reciprocal theorem for Stokes flows, we introduce as a model problem the flow due to a cylinder rotating about its axis with angular velocity $\hat{\bm{\Omega}} = \hat{\Omega} \,\ey$ near a rigid wall at $z=0$. The model velocity field, denoted by $\bvhat(\bm{x})$, satisfies the no-slip conditions $\bvhat = \hat{\bm{\Omega}} \times \bm{r}$ on the cylinder and $\bvhat = \bm{0}$ at $z=0$. 

We have therefore defined both the main and model problems on the undeformed domain. The reciprocal theorem for the two Stokes flows $\bm{v}$ and $\bvhat$ in the same domain and associated, respectively, with stress tensors $\bsigma$ and $\bsigmahat$, states that \citep{leal_book}
\begin{equation} \label{Recip}
	\int_{S_c + S_{w0} + S_{\infty}} \bm{n} \cdot \bsigma \cdot \bvhat \, \rmd S = \int_{S_c + S_{w0} + S_{\infty}} \bm{n} \cdot \bsigmahat \cdot \bv \, \rmd S\,,
\end{equation}
where $S_{\infty}$ is the bounding surface at infinity; here we take $\bm{n}$ as the unit normal pointing into the fluid. Applying the boundary conditions for $\bm{v}$ and $\hat{\bm{v}}$ and noting that the integral over $S_{\infty}$ is vanishingly small, \eqref{Recip} yields
\begin{align}  \label{RecipIntermediate}
	\hat{\bm{\Omega}}~ \cdot \int_{S_c} \bm{r} \times (\bm{n} \cdot \bsigma)\, \rmd S = \bm{\Omega} ~\cdot \int_{S_c} \bm{r} \times (\bm{n} \cdot \bsigmahat) \,\rmd S  + \int_{S_{w0}} \bm{n} \cdot \bsigmahat \cdot \bm{v} \, \rmd S.
\end{align} 
Observing that the integrals over $S_c$ on the left and right sides of the equality in \eqref{RecipIntermediate} are by definition the hydrodynamic torques per unit length on the cylinder $\bm{L}^H = L^H \ey$ and $\hat{\bm{L}}^H = \hat{L}^H \ey$  in the main and model problems, respectively, we find on using \eqref{eq:vbcz0} and \eqref{RecipIntermediate} that 
\begin{equation} \label{eq:recip}
 	\bm{\Omega} \cdot \hat{\bm{L}}^H - \hat{\bm{\Omega}} \cdot \bm{L}^H =  \int_{-\infty}^{\infty} \ez \cdot \bsigmahat \cdot \left\{\bm{v}_c +  \left(u_c \pp{\delta}{x} - \pp{\delta}{t}\right)\ez + \delta \pp{\bm{v}}{z} + \frac{\delta^2}{2} \pp[2]{\bm{v}}{z}\right\} \bigg|_{z=0} \, \rmd x \,.
\end{equation}
If in the main problem, the cylinder is torque-free, \eqref{eq:recip} simplifies to 
\begin{equation} \label{OmegaReciprocal}
\Omega =  \frac{1}{\hat{L}^H}\int_{-\infty}^{\infty} \ez \cdot \bsigmahat \cdot \left\{\bm{v}_c +  \left(u_c \pp{\delta}{x} - \pp{\delta}{t}\right)\ez + \delta \pp{\bm{v}}{z} + \frac{\delta^2}{2} \pp[2]{\bm{v}}{z}\right\} \bigg|_{z=0} \, \rmd x\,. 
\end{equation}
The above expression is generally valid for (i) any type of small surface deformation $\delta$ not limited to the form in \eqref{delta}, and (ii) arbitrary values of $h_f$ relative to $a$. For the remainder of the analysis, however, we focus on deformations given by \eqref{delta} and the lubrication limit $h_f \ll a$. {\color{black} In the sections immediately following, we summarize known properties of the main and the model problems before applying the reciprocal relation \eqref{OmegaReciprocal} to compute the angular velocity.}

\subsection{Model problem: cylinder rotating near a rigid wall} \label{sec:recip}
We summarize here the well-known solution to the model flow driven by a cylinder rotating near a rigid plane wall with angular velocity $\hat{\Omega}$ under the lubrication approximation $h_f \ll a$ \citep{jef81}. On introducing the dimensionless variables
\begin{equation} \label{LubModel}
\begin{array}{ll}
	\m X = \frac{x}{\ell}, \quad Z = \frac{z}{h_f}, \quad \hat{U} = \frac{\hat{u}}{a \hat{\Omega}}, \quad \hat{W} = \frac{\hat{w}}{a \hat{\Omega} h_f/\ell}, 
	\quad \hat{P} =  \frac{\hat{p}}{\mu a \hat{\Omega} \ell/h_f^2} \quad \mbox{and} \quad \ell = \sqrt{2 a h_f},
\end{array}
\end{equation}
the shape of the cylinder in the lubrication layer is locally described by the parabolic profile $Z = H(X) = 1 +X^2$, where $X=0$ represents the position along the wall closest to the surface of the cylinder. The solution to the model problem is 
\begin{subequations}  \label{model}
\begin{align}
	\quad \hat{P}(X) = \frac{2X}{(1 + X^2)^2} \quad \mbox{and} \quad \hat{U}(X,Z) = \frac{1}{2} \dd{\hat{P}}{X} Z (Z-H) - \frac{Z}{H}\,, \tag{\ref{model}a,b}
\end{align} 
\end{subequations}
from which follow
\begin{subequations}  \label{dudz_model}
\begin{equation}
	\pp{\hat{U}}{Z}\bigg|_{Z=0} = \frac{2(X^2-1)}{(1 + X^2)^2} \quad\mbox{and}\quad \pp{\hat{U}}{Z}\bigg|_{Z=H} = -\frac{4 X^2}{(1 + X^2)^2}\,,\tag{\ref{dudz_model}a,b}
\end{equation}
\end{subequations}
which are useful in subsequent calculations. The dimensional surface traction $\ez \cdot \hat{\bm{\sigma}}$ at the wall, which appears in \eqref{OmegaReciprocal}, is therefore 
\begin{equation} \label{ezsigma}
	\ez \cdot \hat{\bm{\sigma}}|_{z=0} = \frac{\mu a \hat{\Omega} \ell}{h_f^2} \left(-\hat{P} \,\ez + \frac{h_f}{\ell}\pp{\hat{U}}{Z}\ex\right)\Bigg|_{Z = 0} .
\end{equation}
The dimensional hydrodynamic torque per unit length of the cylinder is then
\begin{equation}
	\hat{L}^H = \frac{\mu \hat{\Omega} a^2 \ell }{h_f} \int_{-\infty}^{\infty} \pp{\hat{U}}{Z}\bigg|_{Z=H} \rmd X = -\frac{2\sqrt{2} \pi \mu \hat{\Omega} a^{5/2}}{h_f^{1/2}}\,,
\end{equation}
where the negative sign indicates that the hydrodynamic torque opposes the rotation \citep{jef81}. 

\subsection{Torque-free rotation of a cylinder sliding parallel to a soft wall} \label{torquefreecyl}
We now address the question of whether a torque-free cylinder (i.e. $\bm{L}^H = \bm{0}$) translating parallel to a soft wall at constant sliding speed $u_c$ and constant separation distance $h_f$ (i.e. $w_c = 0$) rotates at a finite angular velocity. In this case, the elastic deformation is time-independent in the translating reference frame and can be written as $\delta  = \delta(x)$. This situation is relevant to the steady-state motion of the cylinder free-falling near a soft incline under gravity \citep{sal15}, which we discuss in greater detail in subsequent sections. Before evaluating the integrals in \eqref{OmegaReciprocal} to compute $\Omega$, we give the known solution \citep{sko04,sko05} to the flow up to $\O(\Lambda)$. We define dimensionless flow variables for the main problem (similar to definitions \eqref{LubModel} relevant to the model problem) as
\begin{equation} \label{LubMain}
	P = \frac{p}{\mu u_c \ell/h_f^2}, \quad U = \frac{u}{u_c},\quad\mbox{and}\quad W = \frac{w}{u_c h_f/\ell}\,,
\end{equation}
and introduce a dimensionless velocity vector 
\begin{equation} \label{Vnondim}
	\bm{V}(X,Z) = \frac{\bm{v}}{u_c} = U(X,Z)\, \ex + \frac{h_f}{\ell} W(X,Z) \,\ez\,.
\end{equation}
The dimensionless deformation of the wall can be written using \eqref{delta} and \eqref{LambdaDef} as
\begin{equation} \label{eq:Delta}
	\Delta(X) \equiv \frac{\delta(x)}{h_f} = -\Lambda P(X)\,,
\end{equation}
where $\Lambda$, defined in \eqref{LambdaDef}, is the dimensionless compliance of the soft material, and we have used the simplification that the normal stress is dominated by pressure in the lubrication limit. The solution to the flow problem up to $\O(\Lambda)$ is known analytically \citep{sko05} and can be written using a perturbation expansion as 
\begin{equation} \label{LambdaLubExpansion}
	(P,U,W) = (P_0,U_0,W_0) + \Lambda (P_1,U_1,W_1) + \O(\Lambda^2),
\end{equation}
where
\begin{equation} \label{eq:plub}
	P_0(X) = \frac{2X}{(1+ X^2)^2} \quad \mbox{and} \quad P_1(X) = -\frac{3 (5 X^2 - 3)}{5(1+X^2)^5}.
\end{equation}
The expression for $U(X,Z)$ is
\begin{equation}
	U(X,Z) = \frac{1}{2} \dd{P}{X} (Z-H)(Z+ \Lambda P) - \frac{H - Z}{H+ \Lambda P},	
\end{equation}
which can be expanded in powers of $\Lambda$ to obtain $U_0(X,Z)$ and $U_1(X,Z)$. The following results will be required for subsequent calculations: 
\begin{subequations}  \label{dudz_main}
	\begin{equation}
	\pp{U_0}{Z}\bigg|_{Z=0} = \frac{4 X^2}{(1+X^2)^2}, \quad \pp{U_1}{Z} \bigg|_{Z=0} = - \frac{4 X(5X^2- 3)}{(1+X^2)^5}, \quad U_1\bigg|_{Z=0} = \frac{8 X^3}{(1+X^2)^4}.	\tag{\ref{dudz_main}a,b,c}
	\end{equation}
\end{subequations}

{\color{black} Now that we have laid out the (already known) properties of the main and model problems, it is straightforward to compute $\Omega$ using the reciprocal identity \eqref{OmegaReciprocal}.} We express both $\bm{V}(X,Y)$ and $\Delta(X)$ as series expansions in powers of $\Lambda$:
\begin{subequations}  \label{vdeltaexpansions}
\begin{align} 
	\bm{V} &= \bm{V}_0 + \Lambda \bm{V}_1 + \O(\Lambda^2) \quad \mbox{and} \\
	\Delta &= -\Lambda P_0 - \Lambda^2 P_1 + \O(\Lambda^3),
\end{align}
\end{subequations}
where $\bm{V}_i = U_i \ex + W_i (h_f/\ell) \ez$ with $i \in \{0,1\}$, cf. (\ref{Vnondim}, \ref{LambdaLubExpansion}). Substituting \eqref{ezsigma}, \eqref{Vnondim} and \eqref{vdeltaexpansions} into the reciprocal relation \eqref{OmegaReciprocal}, and collecting powers of $\Lambda$, we can write
\begin{equation} \label{eq:OmegaExpansion}
	\Omega  = \Omega_0 + \Lambda \Omega_1 + \Lambda^2 \Omega_2 + \O(\Lambda^3)\,, 
\end{equation}
where 
\begin{subequations} \label{Omega012}
\begin{align}
	\Omega_0 &= -\frac{u_c}{2 \pi a} \intinf \pp{\hat{U}}{Z}\bigg|_{Z=0} \rmd X\,,\label{Omega0integral}\\
	\Omega_1 &= -\frac{u_c}{2 \pi a} \intinf \left\{ \hat{P} \pp{P_0}{X}  - P_0 \pp{\hat{U}}{Z} \pp{U_0}{Z} \right\}\Bigg|_{Z=0}\rmd X,\quad\mbox{and} \label{Omega1integral}\\
	\Omega_2 &= -\frac{u_c}{2 \pi a} \intinf \left\{\hat{P} \pp{P_1}{X}  - P_0 \left(\pp{\hat{U}}{Z} \pp{U_1}{Z} + \hat{P} \pp{U_1}{X} \right) - P_1 \pp{\hat{U}}{Z} \pp{U_0}{Z} + \frac{P_0^2}{2} \left( \pp{\hat{U}}{Z}\pp{P_0}{X} + \hat{P} \frac{\partial^2 U_0}{\partial X\partial Z} \right) \right\}\Bigg|_{Z=0} \rmd X,\label{Omega2integral} 
\end{align}
\end{subequations}
and we have used $\partial W/\partial Z = - \partial U/\partial X$ and $\partial^2 U_0/\partial Z^2 = \partial P_0/\partial X$.
\begin{comment}
\begin{subequations} \label{Omega012}
\begin{align}
	\Omega_0 &= -\frac{h_f}{2 \pi \mu \hat{\Omega} a^2 \ell} \intinf u_c \,\hatsig_{zx}|_{z=0}\, \rmd x\,, \label{Omega0integral}\\
	\Omega_1 &= -\frac{h_f}{2 \pi \mu \hat{\Omega} a^2 \ell}\intinf \left\{u_c\, \hatsig_{zz} \pp{\delta_0}{x}  + \delta_0\,\ez \cdot \bsigmahat \cdot \pp{\bm{v}_0}{z} \right\}_{z=0} \rmd x\,,\quad\mbox{and} \label{Omega1integral}\\
	\Omega_2 &= -\frac{h_f}{2 \pi \mu \hat{\Omega} a^2 \ell}\intinf\left\{u_c\, \hatsig_{zz} \pp{\delta_1}{x} + \delta_0\,\ez \cdot \bsigmahat \cdot \pp{\bm{v}_1}{z} + \delta_1\,\ez \cdot \bsigmahat \cdot \pp{\bm{v}_0}{z} + \frac{\delta_0^2}{2}\, \ez \cdot \bsigmahat \cdot \pp[2]{\bm{v}_0}{z} \right\}_{z=0}\rmd x\,.
\end{align}
\end{subequations}
\end{comment}
Note that while the lubrication solutions are formally valid within the gap where $|X| = \O(1)$, it is asymptotically accurate to use these solutions over the range $-\infty < X <\infty$, due to the sufficiently rapid decay of the solutions for $X \gg 1$ \citep{sto05b,jef81,leal_book}. {\color{black} Recalling that $\delta/h_f = \O(\Lambda)$, $\bm{v} = \O(u_c)$ and $z = \O(h_f)$, we note that the retention of terms up to $\O(\Lambda^2)$ in \eqref{eq:OmegaExpansion} is entirely consistent with the quadratic order of approximation employed in the domain perturbation expansion \eqref{eq:vbcz0}.}

Substituting (\ref{dudz_model}a) into \eqref{Omega1integral} results in
\begin{equation} \label{Omega0}
	\Omega_0 = -\frac{u_c}{2\pi a} \intinf \frac{2(X^2-1)}{(1 + X^2)^2} \rmd X = 0,
\end{equation}
which corresponds to the nontrivial but well-known result that a torque-free infinite cylinder translating parallel to a rigid wall does not rotate in the absence of inertia \citep{jef81}. Next, the integrand in \eqref{Omega1integral} is an odd function of $X$, guaranteeing that
\begin{equation} \label{Omega1}
	\Omega_1 = 0\,,
\end{equation}
which is in agreement with the result of \citet{sal15} that there is no steady-state rotation of an infinite cylinder at $\O(\Lambda)$. At $\O(\Lambda^2)$, upon substituting \eqref{model}, \eqref{dudz_model}, \eqref{eq:plub} and \eqref{dudz_main} into (\ref{Omega012}c), we find that
\begin{align} \label{Omega2}
	\Omega_2 &= \frac{21}{256} \frac{u_c}{a}.
\end{align}
Substituting (\ref{Omega0})--(\ref{Omega2}) in \eqref{eq:OmegaExpansion} results in a prediction for the rotation rate
\begin{equation} \label{Omegaresult}
	\frac{a \Omega}{u_c} =  \frac{21}{256} \Lambda^2 = \frac{21}{128}\frac{\mu^2 u_c^2 a h_e^2}{h_f^5 (2G + \lambda)^2}\,,
\end{equation}
placing the scaling relation \eqref{scaling} on quantitative footing. The softness of the wall therefore causes the cylinder to rotate with an angular velocity that is quadratic in the dimensionless compliance $\Lambda$. {\color{black} The sense of rotation is the same as that of frictional (solid-solid) rolling without slip}. 

The above result can also be stated in terms of the elastohydrodynamic torque $\bm{L}^H = L^H \ey$ induced on a cylinder that slides without rotation. Setting $\bm{\Omega} = \bm{0}$ in \eqref{eq:recip}, we find
\begin{equation}
	L^H_{\bm{\Omega} = \bm{0}} = - \frac{21}{256}  \frac{2 \pi \mu u_c a \sqrt{2 a h_f}}{h_f} \Lambda^2  = -\frac{21 \sqrt{2} \pi}{64} \frac{\mu^3 u_c^3 a^{5/2} h_e^2}{h_f^{11/2} (2G + \lambda)^2}\,.
\end{equation}
which is quadratic in $\Lambda$ and consequently depends on the cube of the sliding speed. By contrast, the elastohydrodynamic lift force on the cylinder depends linearly on $\Lambda$ for $\Lambda \ll 1$ \citep{sko05,sal15}.

\begin{figure}
\includegraphics[scale=1]{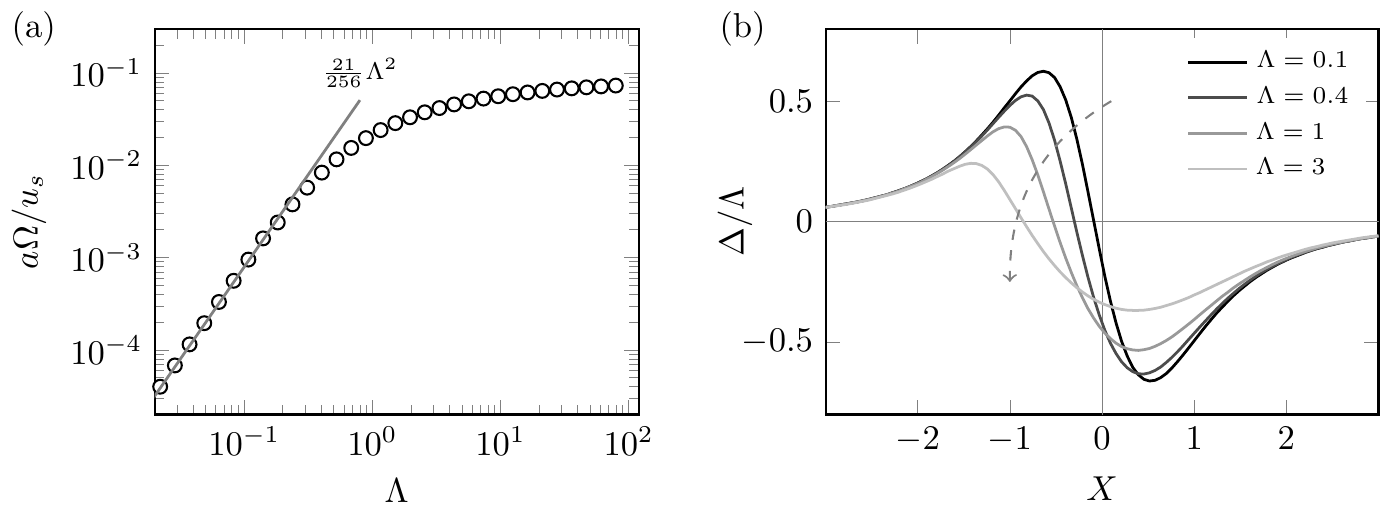}
\caption{(a) Dimensionless rotation rate $a\Omega/u_c$ at zero external torque as a function of the dimensionless compliance $\Lambda$, showing the theoretical prediction \eqref{Omega2} for $\Lambda \ll 1$ and numerical solutions (symbols) of the system \eqref{LubEqnExact}, \eqref{ZeroTorqueCondition}. (b) Rescaled dimensionless deformation of the wall (equal to the negative pressure field, $\Delta(X)/\Lambda = -P(X)$) obtained from a numerical solution of the nonlinear lubrication equation \eqref{LubEqnExact} subject to the zero-torque condition \eqref{ZeroTorqueCondition}. Curves correspond to $\Lambda = 0.1$, 0.4, 1 and 3 with the dashed arrow indicating the direction of increasing $\Lambda$.}
\label{FigSimvTheory}
\end{figure}

{\color{black} The asymptotic result \eqref{Omegaresult} is verified against a numerical solution of the Reynolds equation for the fluid flow in the lubrication layer (derived in Appendix \ref{app})
\begin{equation}\label{LubEqnExact}
	\dd{}{X}\left((H+\Lambda P)^3\dd{P}{X} + 6(H+\Lambda P)(1+\beta) \right) = 0,
\end{equation}
subject to the boundary condition $P(X \rightarrow \pm \infty) \rightarrow 0$, where $P(X)$ is the dimensionless pressure, $\beta \equiv a \Omega/u_c$ is the dimensionless rotation rate of the cylinder and we recall that $H(X) = 1+X^2$ is the dimensionless shape of the cylinder's surface in the lubrication region. The value of $\beta$ is determined as a part of the solution by the constraint of zero hydrodynamic torque on the cylinder (Appendix \ref{app})
\begin{equation} \label{ZeroTorqueCondition}
	0 = \int_{-\infty}^{\infty} \pp{U}{Z}\bigg|_{Z=H} \rmd X = \int_{-\infty}^{\infty} \frac{\frac{1}{2}\dd{P}{X} (H + \Lambda P)^2 + 1-\beta}{(H+\Lambda P)} \rmd X\,.
\end{equation}
The differential equation \eqref{LubEqnExact} for $P(X)$ is solved numerically using the \emph{Matlab} routine {\em bvp5c}, in conjunction with a shooting method to self-consistently determine the value of $\beta$ that satisfies the zero-torque condition \eqref{ZeroTorqueCondition}. The numerical results for the zero-torque rotation rate are plotted as a function of $\Lambda$ in figure \ref{FigSimvTheory}(a), confirming the asymptotic result \eqref{Omegaresult} for $\Lambda \ll 1$. Numerical results for the dimensionless deformation of the wall $\Delta(X) = -\Lambda P(X)$ are shown in figure \ref{FigSimvTheory}(b). We also compute results for the drag and lift forces on the cylinder as functions of $\Lambda$, as discussed in appendices \ref{app} and \ref{app:LiftForce}. For the remainder of this article, we will restrict our attention to the small-$\Lambda$ asymptotic limit \eqref{Omegaresult}.
}

\subsection{Buoyancy-driven sliding down a soft-coated inclined wall} \label{sec:gravity}

\begin{figure}[t]
	\centering
	\includegraphics[scale=1]{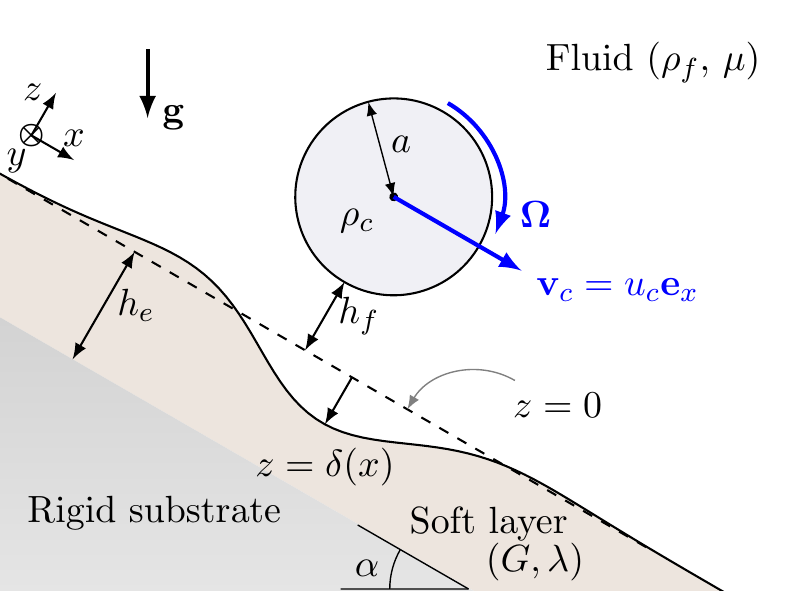}
	\caption{Sketch of a cylinder of density $\rho_f$ sliding parallel to a soft-coated inclined wall under gravity. The balance of hydrodynamic forces and the buoyancy of the cylinder maintains a constant sliding speed $u_c$ and a lubrication layer thickness $h_f$. The flow is steady in the frame of reference of the sliding cylinder.}
	\label{SetupGravity}
\end{figure}

We now apply the theoretical result \eqref{Omegaresult} to the situation in which an immersed cylinder translates parallel to a thin soft coating on a wall that makes an angle $\alpha$ with the horizontal (gravity acts vertically), as sketched in figure \ref{SetupGravity}. Here, we consider a cylinder of density $\rho_c > \rho_f$ and define $\rho^* = \rho_c - \rho_f >0$. At steady state, the translation of the cylinder is determined by a balance between the net buoyant force on the cylinder and hydrodynamic forces. In particular, the hydrodynamic drag on the cylinder balances its net weight tangent to the incline ($\rho^* g a^2 \sin \alpha \sim \mu u_c \ell/h_f$) and the elastohydrodynamic lift force balances the cylinder's net weight normal to the incline ($\rho^* g a^2 \cos \alpha \sim \Lambda \mu u_c \ell^2/h_f^2 $) \citep{sal15,sai16}, where $\ell = \sqrt{2 a h_f}$. At steady state, this balance of forces establishes a constant speed $u_c$ and a constant lubrication gap width $h_f$, which are given by
\begin{subequations} \label{sliding_gravity}
	\begin{align}
	&u_c = A \frac{\rho^* g a^2 \sin \alpha}{\mu} \left(\frac{\rho^* g h_e \cos \alpha}{2G + \lambda} \right)^{1/5}\left(\tan \alpha\right)^{2/5} \quad \mbox{and}\\
	&h_f =  B a \left(\frac{\rho^* g h_e \cos \alpha}{2G + \lambda} \right)^{2/5}\left(\tan \alpha\right)^{4/5},
	\end{align} 
\end{subequations}
where $A = \frac{3^{1/5}}{2^{12/5}}$ and $B = \frac{3^{2/5}}{2^{9/5}}$ are numerical constants that are obtained from a detailed hydrodynamic calculation \citep{sal15}. Substituting the expressions (\ref{sliding_gravity}a,b) for $u_c$ and $h_f$ into \eqref{LambdaDef}, the dimensionless compliance for buoyancy-driven motion is 
\begin{equation}
\Lambda = \frac{\sqrt{2} A}{B^{5/2}} \left(\frac{\rho^* g h_e \cos \alpha}{2G + \lambda} \right)^{1/5}\left(\tan \alpha\right)^{-3/5}\,.
\label{eq:lambdaIncline}
\end{equation}

Since the cylinder is free of external torque, the above expressions can be substituted into \eqref{Omegaresult} to obtain
\begin{align} \label{OmegaGravity}
\frac{a \Omega}{u_c}  = \frac{21}{128} \frac{A^2}{B^5} \left(\frac{\rho^* g h_e \cos \alpha}{2G + \lambda} \right)^{2/5}\left(\tan \alpha\right)^{-6/5},
\end{align}
or
\begin{equation}\label{OmegaGravityDim}
\Omega  = \frac{21}{128} \frac{A^3}{B^5} \frac{\rho^* g a \sin \alpha}{\mu} \left(\frac{\rho^* g h_e \cos \alpha}{2G+\lambda} \right)^{3/5}\left(\tan \alpha\right)^{-4/5}.
\end{equation}

{\color{black} The above results for buoyancy-driven motion are applicable in the limit of small $\Lambda$. We note that the theoretical predictions in sections \ref{sec:Recip} and \ref{sec:gravity} (and the corresponding numerical results) utilize the Winkler approximation \eqref{delta}, wherein $\lambda/G = \O(1)$ is assumed. The limit of an incompressible layer ($\lambda/G \rightarrow \infty$) must generally be treated separately; we develop scaling laws for this case in section \ref{sec:Incomp}.}

\section{Discussion}
{\color{black} 
We have thus shown that a cylinder that translates parallel to a soft wall in a viscous fluid under the action of an external force must also rotate if no external torque acts on it. Within the lubrication limit and the elastic model used here, the rotation rate is determined by a single dimensionless ratio $\Lambda$ that measures the magnitude of the elastic deformation amplitude relative to the fluid gap thickness. We find that the dimensionless rotation rate approaches zero with increasing modulus ($2G + \lambda$) of the elastic layer, provided that all other physical parameters involved in $\Lambda$, in particular the translational speed $u_c $ and the gap height $h_f$ remain unchanged.  The numerical results of figure \ref{FigSimvTheory}(a) show that the induced rotation is relatively weak ($a \Omega/u_c \lesssim 0.1$ for $\Lambda \lesssim 100$). Also, the rotation rate approaches zero as $\Lambda \rightarrow 0$, consistent with previously known results for the rotation of infinite cylinders \citep{jef81,sal15}}.

For a cylinder that is driven through a viscous fluid along a soft incline by gravity, the translational speed and the lubrication layer thickness are simultaneously determined by force balances. These balances self-consistently set the dimensionless compliance $\Lambda$, cf. \eqref{eq:lambdaIncline}. For small deformations $(\Lambda \ll 1)$, the translational speed $u_c$, the thickness of the layer $h_f$ and the angular velocity $\Omega$ all increase with the softness of layer (decreasing $2G + \lambda$). 

\subsection{General remarks on symmetry} \label{sec:symmetry}
We briefly discuss some general geometrical features of the problem for wall-parallel steady translation of a torque-free cylinder ($w_c =0, \bm{L}^H = \bm{0})$. First, we note that for translation parallel to a rigid wall, a torque-free cylinder does not rotate at all. This is a nontrivial but well known result that applies only to infinite cylinders, but not more generally to compact objects such as spheres \citep{jef81,gol67a,gol67b}. 

Translation of the cylinder parallel to the wall drives a flow through the gap, which at $\O(\Lambda^0)$ corresponds to an antisymmetric pressure distribution about the symmetry plane $X=0$. This pressure distribution results in an antisymmetric deformation of the interface at $\O(\Lambda)$ \citep{sko04}. The $\O(\Lambda)$ flow has the opposite symmetry, i.e. it corresponds to a symmetric pressure distribution $P_1(X)$, which is ultimately responsible for the elastohydrodynamic lift force exerted on the cylinder. By the same token, the shear stress in the fluid at $\O(\Lambda)$ has the opposite (odd) symmetry about $X=0$ and therefore cannot produce any torque on the cylinder.

At next order, the symmetries of the pressure and shear stresses become reversed yet again. The pressure of the $\O(\Lambda)$ flow creates a symmetric deformation of the wall at $\O(\Lambda^2)$. The corresponding $\O(\Lambda^2)$ flow is characterized by an antisymmetric pressure distribution $P_2(X)$, and a symmetric shear stress distribution on the surface of the cylinder, about the $X=0$ plane. Thus, the $\O(\Lambda^2)$ problem realizes the symmetry that is required to produce an elastohydrodynamic torque on a purely sliding cylinder. We remark that additional corrections to the drag are expected at $\O(\Lambda^2)$ due to the same reasons of symmetry, but no additional lift force is produced on the cylinder at this order. The full solution to the $\O(\Lambda^2)$ problem is given in Appendix \ref{app}, in which we compute both the drag force correction as well as provide an alternative method to compute the angular velocity $\Omega$. The drag force is found to be of the form 
\begin{align} \label{DragForce}
	F_{\rm drag}^H = -\frac{2 \sqrt{2} \pi \mu u_c a^{1/2}}{h_f^{3/2}} \left(1 - \frac{441}{1280}\Lambda^2 + \O(\Lambda^4)\right).
\end{align}
Applying the reciprocal theorem as in section \ref{sec:Recip}, we also compute the elastohydrodynamic lift force on a sliding torque-free cylinder as (see Appendix \ref{app:LiftForce})
\begin{equation} \label{LiftForce}
F^H_{\mathrm{lift}} = \frac{2 \mu u_c a}{h_f} \left(\frac{3 \pi}{8} \Lambda - \frac{2157\pi}{2048} \Lambda^3 + \O(\Lambda^5)\right)\,.
\end{equation}
The term in \eqref{LiftForce} proportional to $\Lambda$ is the leading-order result that is well known \citep{sek93,sko04}, but is obtained here by the use of the Lorentz reciprocal theorem. The $\O(\Lambda^3)$ correction to the lift force utilizes the solution of the lubrication flow accurate to $\O(\Lambda^2)$, see Appendix \ref{app}. {\color{black} The higher corrections to the lift and drag forces are compared against numerical results in figure \ref{LiftForceFig}.}

As a general rule for a symmetric object translating parallel to a wall in the small-deformation limit, lift forces can only be expected to occur with odd powers of $\Lambda$, and torques and drag forces to occur with even powers of $\Lambda$. The case of the infinite cylinder is further specialized due to the vanishing torque at $\O(\Lambda^0)$, which causes the leading-order torque to be effective at $\O(\Lambda^2)$. By a similar line of argument, the problem in which the cylinder translates perpendicular to the wall ($w_c \neq 0$, $u_c = 0$) has the opposite symmetry.

\begin{comment}
Applying the reciprocal theorem as in section \ref{sec:Recip}, we also compute the elastohydrodynamic lift force on a sliding torque-free cylinder as (see Appendix \ref{app:LiftForce})
\begin{equation} \label{LiftForce}
F^H_{\mathrm{lift}} = \frac{2 \mu u_c a}{h_f} \left(\frac{3 \pi}{8} \Lambda - \frac{2157\pi}{2048} \Lambda^3 + \O(\Lambda^5)\right)\,.
\end{equation}
The term in \eqref{LiftForce} proportional to $\Lambda$ is the leading-order result that is well known \citep{sek93,sko04}, but is obtained here by the use of reciprocal theorem. The $\O(\Lambda^3)$ correction to the lift force utilizes the solution of the lubrication flow accurate to $\O(\Lambda^2)$, which is given in Appendix \ref{app}, where we also calculate the elastohydrodynamic correction to the drag on the cylinder and obtain
\begin{align}
	F_{\rm drag}^H = -\frac{2 \sqrt{2} \pi \mu u_c a^{1/2}}{h_f^{3/2}} \left(1 - \frac{441}{1280}\Lambda^2 + \O(\Lambda^4)\right).
\end{align}
{\color{black} The higher corrections to the lift and drag forces are compared against numerical results in figure \ref{LiftForceFig}.}
\end{comment}

\subsection{Comparison with experiments and open questions}
For the case of buoyancy-driven motion of an immersed cylinder near a soft incline, the predictions of the present theory for the rotation speed (equations \eqref{OmegaGravity} and \eqref{OmegaGravityDim}) are in qualitative agreement with the experimental observation that the rotation rate is typically greater for a soft elastic coating than for a rigid substrate (Supplementary Information of \citet{sai16}). However, the cylinders were also observed to rotate close to a rigid wall in the experiments (albeit very slowly), while the present theory predicts no rotation in this case $(\Lambda = 0)$. Additionally, for finite softness, the experimental rotation rates appear to be somewhat greater than those predicted by figure \ref{FigSimvTheory}(a).

%However, while the two-dimensional theory (and corresponding numerical results) developed here predicts only a weak rotation of the cylinder ($a \Omega/u_c \lesssim 0.1$ up to large values of the compliance $\Lambda \lesssim 100$, see figure \ref{FigSimvTheory}), the rotation rates observed in the experiments are typically greater \citep{sai16}. In addition, the short cylinders used in experiments are observed to rotate even close to a rigid wall, while the 2D theory predicts no rotation of a cylinder near a rigid wall (see supplementary information of \citet{sai16}). 

A likely source of this disagreement is that the theory is strictly applicable in the limit of infinite cylinders ($L/a \rightarrow \infty$), while those used in the experiments have $L/a \lesssim 2$. %The finite aspect ratio of the cylinder produces three-dimensional flows near its ends, which (i) produces an additional torque on the cylinder due to stresses on its side faces, and (ii) modifies the deformation of the elastic layer. 
A finite cylinder can be expected to rotate with non-zero $\Omega$ even as $\Lambda \rightarrow 0$ (the limit of a rigid wall) due to end-effects; a well-known analog is that of a \emph{sphere} translating parallel to a rigid wall, which, in the lubrication limit, must rotate such that $a \Omega/u_c \approx \frac{1}{4}$ \citep{gol67a} in order to remain torque-free. In general, due to the symmetry arguments of section \eqref{sec:symmetry}, we expect the infinite-cylinder theory for rotation \eqref{eq:OmegaExpansion} to become modified for finite cylinders according to 
\begin{equation}
	\Omega = \Omega_0 + \Lambda^2  \Omega_2 + \O(\Lambda^4) \quad \mbox{for} \quad \Lambda \ll 1,
\end{equation}
where $\Omega_0$ and $\Omega_2$ depend on $u_c, a, h_f$ and $L$, and $\Omega_0 \rightarrow 0$ as $L/a \rightarrow \infty$, as in \eqref{Omega0}. From dimensional analysis, we expect $a \Omega_0/u_c = f_0(h_f/a,L/a)$ and $a \Omega_2/u_c = f_2(h_f/a,L/a)$ with $f_0$ and $f_2$ being unknown functions that approach zero and $\frac{21}{256}$, respectively, as $L/a$ approaches infinity, cf. \eqref{Omega0}, \eqref{Omegaresult}. Unfortunately, even the form of $f_0(h_f/a,L/a)$ (the limit of a rigid wall) is not known and requires a numerical treatment of the 3D flow; we are not aware of such a study of a finite-aspect-ratio cylinder translating near a rigid wall. 

We also note that the modeling of the elastic response as linear and local (the Winkler approximation) is formally applicable in the limit where $|\delta|^2/a \ll h_e^2/a \ll h_f$. For buoyancy-driven viscous sliding near an incline, this separation of scales becomes increasingly strained for soft substrates (decreasing $2G + \lambda$) due to a combination of effects. First, the elastic deformation $\delta$ increases with the softness of the substrate. Second, the elastohydrodynamic lift force diminishes with $\Lambda$ for $\Lambda \gtrsim \O(1)$ \citep{sko04}, so the steady-state gap thickness $h_f$ must decrease in order for the lift force to support the normal component of the cylinder's buoyant weight. In this limit of large deformations, a nonlocal elastic description similar to that of \citet{sno13} for heavily loaded elastic layers may be more appropriate. A systematic quantification of steady-state rotation in the experiments, including the influence of the cylinder ends and the effect of large deformations of the substrate, should be discussed in future work.

\subsection{Incompressible elastic layer} 
\label{sec:Incomp}
{\color{black} Retaining the assumption of small deformations and invoking the symmetry arguments of section \ref{sec:symmetry}, we now develop scaling relations for the angular velocity of a cylinder sliding near an incompressible, thin, elastic layer ($\lambda/G \rightarrow \infty)$.}   Incompressibility of the thin elastic layer (we assume $h_e \ll \ell$) requires that $\delta_x'/\ell \sim \delta'/h_e$, where $\delta_x'$ is the scale of elastic deformation in the $x-$direction (parallel to the sliding) and $\delta'$ is the deformation normal to the wall; the prime denotes incompressibility of the elastic layer. The stress balance in the thin elastic layer is lubrication-like, i.e. the pressure in the elastic solid scales as $p_e = \O(G \delta_x' \ell/h_e^2)$. The elastic pressure must balance the pressure in the fluid $p =\O(\mu u_c \ell/h_f^2)$ to keep the fluid-solid interface in quasi-static equilibrium. This balance of pressures $p \sim p_e$ yields \citep{sko04}
\begin{equation}
	\delta' \propto \frac{\mu u_c h_e^3}{h_f^2 \ell G}\,.
\end{equation}
The dimensionless compliance (the elastic deformation amplitude relative to the lubrication layer thickness) is therefore 
\begin{equation}
	\Lambda' \equiv \frac{\mu u_c h_e^3}{a^{1/2} h_f^{7/2} G}\,.
\end{equation}
Following the discussion on symmetry in section \ref{sec:symmetry}, we generically expect
\begin{equation}
	\frac{a \Omega}{u_c} \propto (\Lambda')^2 \sim \frac{\mu^2 u_c^2 h_e^6}{a h_f^7 G^2}\,
\end{equation}
for an infinite cylinder. 

For a cylinder sedimenting due to buoyancy near an incline coated with a thin incompressible elastic layer, the balance of hydrodynamic forces and the cylinder's net weight, assuming small deformations ($\Lambda' \ll 1$) leads to $\rho^* g a^2 \sin \alpha \sim \mu u_c \ell/h_f$ (tangent to the incline) and $\rho^* g a^2 \cos \alpha \sim \Lambda' \mu u_c \ell^2/h_f^2 $ (normal to the incline), now resulting in
\begin{align}
&u_c \propto \left(\frac{\rho g a^2 \sin \alpha}{\mu}\right) \left( \frac{\rho g h_e \cos \alpha}{G} \frac{h_e^2}{a^2}\right)^{1/7}  \left(\tan \alpha\right)^{2/7}, \\ 
&h_f \propto a \left( \frac{\rho g h_e \cos \alpha}{G} \frac{h_e^2}{a^2}\right)^{2/7}  \left(\tan \alpha\right)^{4/7}. 
\end{align}
The rotation rate is then 
\begin{equation}
	\frac{a \Omega}{u_c} \propto \left( \frac{\rho g h_e^3 \cos \alpha}{G a^2}\right)^{2/7} \left(\tan \alpha\right)^{-10/7}\,.
\end{equation}
The dependence of $\Omega$ with $G$ for an incompressible material is weaker $(\Omega \propto G^{-3/7})$ compared with the prediction of the local model in \eqref{delta}, $\Omega \propto G^{-3/5}$, see \eqref{OmegaGravityDim}.

\section{Conclusions}
{\color{black} Using the Lorentz reciprocal theorem for Stokes flows, we have shown in the lubrication limit that an infinite cylinder sliding parallel to a soft wall in a viscous fluid without rotation experiences an elastohydrodynamic torque in addition to the well-known elastohydrodynamic lift force. The induced torque is quadratic in the deformation $\delta$ of the wall, while the lift force scales linearly with deformation. Thus, we find that a cylinder free of external torque sliding parallel to a soft wall must rotate with a non-zero angular velocity in the same sense as expected in rigid-body frictional rolling. The angular velocity $\Omega$ scales as the cube of the sliding speed and the square of the viscosity, and increases with the softness of the elastic layer. The theoretical predictions for the rotation rate in the limit of small elastic compliances are in agreement with numerical solutions for finite compliances within the framework of linear elasticity. 

{The results of the model qualitatively describe previous experiments on an immersed cylinder free-falling and rotating under gravity near an inclined soft-coated wall, namely that soft coatings typically result in greater rotation speeds than rigid walls. However, the present theoretical results appear to underpredict the previously observed rotation rates, likely due to three-dimensional end-effects in the latter.} %The differences may be ascribed to (i) the torque produced by three-dimensional flows near the ends of the finite-length cylinder, and  (ii) deficiencies in the modeling of the soft layer, which may involve a nonlinear, nonlocal or poroelastic response.
The geometric structure of the problem results in symmetry arguments pertaining to the motion of symmetric objects such as cylinders or spheres near weakly deformable boundaries. These general arguments are likely to be useful in hydrodynamically mediated interactions between soft elastic surfaces, which are important features of some biological and geophysical systems.}

\begin{acknowledgments}
We thank the Carbon Mitigation Initiative of Princeton University for partial support of this research. T.S. acknowledges financial support from the Global Station for Soft Matter, a project of Global Institution for Collaborative Research and Education at Hokkaido University. We thank Martin Essink, Anupam Pandey and Jacco Snoeijer for suggesting the possible role of incompressibility in our problem.
\end{acknowledgments}

\appendix 
\section{Direct calculation of $\Omega$ and flow up to $\O(\Lambda^2)$}\label{app}

A direct calculation of the $\O(\Lambda^2)$ flow by solving the lubrication equations provides a {\color{black} conventional} alternative  method to compute $\Omega$. In addition, we also find the $\O(\Lambda^2)$ correction to the drag force on the cylinder. We consider only the case where the cylinder is torque-free and its center translates parallel to the soft wall (i.e. at constant $h_f$) with constant translational speed $u_c$ and rotates at an unknown angular speed $\Omega$. If we define a dimensionless angular speed
\begin{equation}
	\beta \equiv \frac{a \Omega}{u_c}\,,
\end{equation}
and employ the usual lubrication scaling, cf. (\ref{LubModel},\ref{LubMain}), the boundary conditions for the flow $(U,W)$ are
\begin{subequations}
\begin{align}
	&U = -\beta, \quad W = -\beta \dd{H}{X}\quad \mbox{on}\quad  Z = H(X) = 1+X^2\,, \\
	&U = -1, \quad W = -\dd{\Delta}{X} \quad \mbox{on}\quad Z = \Delta(X),
\end{align}
\end{subequations} 
where $\Delta \equiv \delta/h_f$. The momentum equations in the lubrication limit reduce to $P = P(X)$ and 
\begin{equation}
	U = \frac{1}{2}\dd{P}{X} (Z-H)(Z-\Delta) - \frac{H-Z}{H-\Delta} - \beta\frac{Z-\Delta}{H-\Delta}.
\end{equation}
Integrating the continuity equation between $Z=\Delta$ and $Z=H$, and applying the boundary conditions on $W$ yields the following ordinary differential equation for $P$:
\begin{equation} \label{LubEqnExact2}
	\dd{}{X}\left((H+\Lambda P)^3\dd{P}{X} + 6(H+\Lambda P)(1+\beta)\right) = 0,
\end{equation}
where we have used $\Delta = -\Lambda P$, as per \eqref{eq:Delta}. The dimensionless rotation rate $\beta$ is unknown and must be determined as a part of the solution using the condition of zero torque on the cylinder which, in the lubrication limit, can be written as
\begin{equation}
	0 = \int_{-\infty}^{\infty} \pp{U}{Z}\bigg|_{Z=H} \rmd X = \int_{-\infty}^{\infty} \frac{\frac{1}{2}\dd{P}{X} (H + \Lambda P)^2 + 1-\beta}{(H+\Lambda P)} \rmd X\,.
\end{equation}
The above system can be solved subject to the boundary conditions  $P \rightarrow 0$ as $X \rightarrow \pm \infty$. 

We develop a perturbation solution of the form 
\begin{equation}
	(P,U,W,\beta) = (P_0,U_0,W_0,\beta_0) + \Lambda(P_1,U_1,W_1,\beta_1) + \Lambda^2(P_2,U_2,W_2,\beta_2) + \dots\,.
\end{equation}
Note that we have also expanded $\beta$, since it is determined automatically by the condition of zero torque. At $\O(\Lambda^0)$, we find from \eqref{LubEqnExact2} that 
\begin{equation}
	\dd{}{X}\left((1+X^2)^3\dd{P_0}{X} + 6(1+X^2)(1+\beta_0)\right) = 0,
\end{equation}
which has the solution
\begin{equation}
	P_0(X) = \frac{2X}{(1+X^2)^2} + \frac{2X \beta_0}{(1+X^2)^2}\implies \pp{U_0}{Z}\bigg|_{Z=H} = -\frac{2(X^2-1)}{(1+X^2)^2} - \frac{4 X^2 \beta_0}{(1+X^2)^2}.
\end{equation}
The zero-torque condition then yields
\begin{equation}
\int_{-\infty}^{\infty} \pp{U_0}{Z}\bigg|_{Z=H} \rmd X = -2\upi \beta_0 =  0 \iff \beta_0 = 0.
\end{equation}

At $\O(\Lambda)$, it follows from \eqref{LubEqnExact2} that the equation for the pressure is 
\begin{equation}
	\dd{}{X}\left((1+X^2)^3\dd{P_1}{X} + 3 (1+X^2)^2 P_0 \dd{P_0}{X} + 6\left(\beta_1 (1+X^2) + P_0\right)\right) = 0,
\end{equation}
which has the solution 
\begin{equation}
	P_1(X) = -\frac{3(5X^2 - 3)}{5(1+X^2)^5} + \frac{2X \beta_1}{(1+X^2)^2} \implies \pp{U_1}{Z}\bigg|_{Z=H} = \frac{4X(X^2-3)}{(1+X^2)^5} -  \frac{4 X^2 \beta_1}{(1+X^2)^2}.
\end{equation}
The zero-torque condition again results in
\begin{equation}
	\int_{-\infty}^{\infty} \pp{U_1}{Z}\bigg|_{Z=H} \rmd X = -2\upi \beta_1 {=} 0 \iff \beta_1 = 0,
\end{equation}
as recognized by \citet{sal15}. The equation for the pressure at $\O(\Lambda^2)$ is 
\begin{equation}
	\dd{}{X}\left((1+X^2)^3\dd{P_2}{X} + 3 (1+X^2)^2 \left(P_0 \dd{P_1}{X} + P_1 \dd{P_0}{X}\right) + 3(1+X^2)P_0^2 \dd{P_0}{X} + 6\left(\beta_2 (1+X^2) + P_1\right)\right) = 0,
\end{equation}
which has the solution (obtained using \emph{Mathematica})
\begin{align}
	P_2 = -\frac{X \left(6699 X^{10}+39237 X^8+95062 X^6+121394 X^4+38815 X^2+87465\right)}{5600 \left(X^2+1\right)^8} + \frac{2X \beta_2}{(1+X^2)^2} \nonumber \\
	\implies \pp{U_2}{Z}\bigg|_{Z=H} = \frac{957 \left(X^2+1\right)^6-4480 \left(X^2+1\right)^2+28672 \left(X^2+1\right)-27648}{320 \left(X^2+1\right)^8}-\frac{4 X^2 \beta_2}{\left(X^2+1\right)^2}\,.
\end{align}
The zero-torque condition now results in
\begin{equation}
\int_{-\infty}^{\infty} \pp{U_2}{Z}\bigg|_{Z=H} \rmd X = 2\upi\left(\frac{21}{256} - \beta_2\right) = 0 \iff \beta_2 = \frac{21}{256}.
\end{equation}
Thus, we find that a torque-free infinite cylinder sliding parallel to a soft plane has a rotation speed that is given at leading order by
\begin{equation}
	\frac{a \Omega}{u_c} = \frac{21}{256} \Lambda^2,
\end{equation}
which is identical to the result obtained in \eqref{Omegaresult} using the reciprocal theorem. 

\begin{figure}
	\centering
	\includegraphics[scale=1]{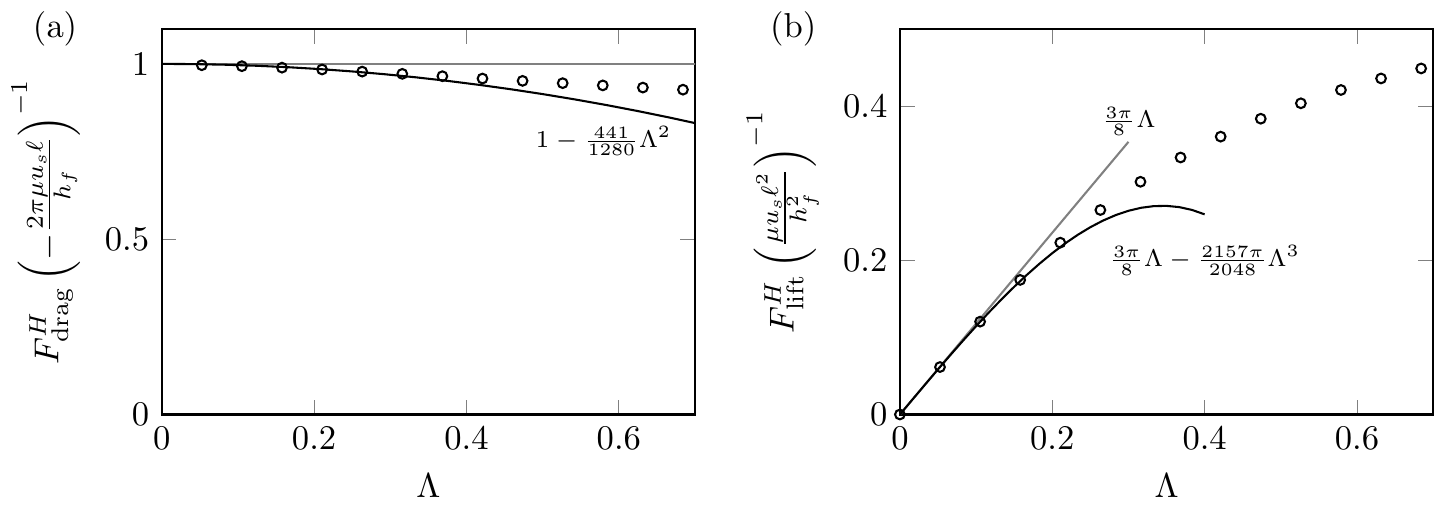}
	\caption{Rescaled (a) drag force and (b) lift force on the cylinder per unit length, showing numerical results (symbols) and theoretical predictions of \eqref{DragForce} and \eqref{LiftForce} (curves). The drag force decays monotonically (but weakly) with $\Lambda$, while the lift force is found to have a maximum at $\Lambda \approx 2.2$.}
	\label{LiftForceFig} 
\end{figure}

We also compute the $\O(\Lambda^2)$ correction to the hydrodynamic drag force. The hydrodynamic drag force per unit length of the cylinder, is given by $F_{\mathrm{drag}}^H = \int_{S_c} \bm{n} \cdot \bm{\sigma} \cdot \ex \rmd x$, where $\bm{n}$ is the normal to the surface of the cylinder directed into the fluid. In the lubrication limit, the above expression reduces to 
\begin{align}
 F_{\rm drag}^H = -\frac{\mu u_c \sqrt{2 a h_f}}{h_f^2} \intinf \left(\pp{U}{Z} + 2 X P\right)_{Z = H} \rmd X.
\end{align}
Using the expansions of $U$ and $P$ in powers of $\Lambda$, we obtain 
\begin{align} \label{FdragApp}
	F_{\rm drag}^H = -\frac{2 \sqrt{2} \pi \mu u_c a^{1/2}}{h_f^{1/2}} \left(1 - \frac{441}{1280}\Lambda^2 + \O(\Lambda^4)\right).
\end{align}
{\color{black} Note that $F_{\rm drag}^H$ is independent of the value of $\beta_2$. The elasticity of the layer lowers the drag on the cylinder, although this is a weak effect for modest values of $\Lambda$. This is evident in figure \ref{LiftForceFig}(a), where the asymptotic form of $F_{\rm drag}^H$ is plotted alongside numerical calculations at finite $\Lambda$.}

The perturbation analysis may, in principle, be carried forward to higher order in $\Lambda$, although the solutions for the pressure distribution $P(X)$ become increasingly unwieldy with increasing powers of $\Lambda$. However, having developed the solution to the problem up to $\O(\Lambda^2)$, we can now apply the reciprocal theorem again to infer some properties of the flow at $\O(\Lambda^3)$. In particular, we compute the correction to the leading order lift force prediction in Appendix \ref{app:LiftForce}, for which the following results are useful:
\begin{subequations}
	\begin{align}
	&U_2\big|_{Z=0} = \frac{8 X^2 \left(17-25 X^2\right)}{5 \left(X^2+1\right)^7}, \quad \mbox{and} \\
	&\pp{U_2}{Z}\bigg|_{Z=0} = -\frac{957 X^{12}+5742 X^{10}+14355 X^8+19140 X^6-15725 X^4+42862 X^2-2499}{320 \left(X^2+1\right)^8} +  \frac{2 \beta_2  \left(X^2-1\right)}{\left(X^2+1\right)^2}\,.
	\end{align}
\end{subequations}

\section{Lift force at $\O(\Lambda^3)$ using the reciprocal theorem}\label{app:LiftForce}
To compute the lift force up to $\O(\Lambda^3)$ using the reciprocal theorem, we define a model problem in which a cylinder translates \emph{normal} to a rigid wall with velocity $\tilde{\bm{V}} = \tilde{V} \ez$, thereby driving a model flow $(\tilde{\bm{v}}, \tilde{\bm{\sigma}})$. We now apply the reciprocal theorem in the undeformed domain in a manner analogous to that in \eqref{sec:Recip}, using the new model problem. Doing so results in the following expression for the normal force per unit length of the cylinder: 
\begin{align}\label{ForceRecip}
\bm{F}^H \cdot \tilde{\bm{V}} = -\int_{\infty}^{\infty}\ez \cdot \tilde{\bsigma} \cdot \left\{\bm{v}_c +  \left(u_c \pp{\delta}{x} - \pp{\delta}{t}\right)\ez + \delta \pp{\bm{v}}{z} + \frac{\delta^2}{2} \pp[2]{\bm{v}}{z} + \frac{\delta^3}{6} \pp[3]{\bm{v}}{z}\right\} \bigg|_{z=0} \, \rmd x + \O\left(\left|\delta^4 \ez \cdot \tilde{\bsigma} \cdot \pp[4]{\bm{v}}{z}\right|\right)\,,
\end{align}
where the boundary condition at $z=0$ in the main problem has now been expanded to one power of $\delta$ higher than in \eqref{eq:vbcz0}. The solution to the model flow is well known in the lubrication limit \citep{jef81}; the surface traction on the wall in the model problem, which enters \eqref{ForceRecip} is 
\begin{equation}
\ez \cdot \tilde{\bsigma}|_{z=0} = \frac{\mu \tilde{V} \ell^2}{h_f^3} \left(- \tilde{P} \ez + \frac{h_f}{\ell}\pp{\tilde{U}}{Z} \ex\right)\Bigg|_{Z=0}, \quad \mbox{where}\quad  \tilde{P} = -\frac{3}{(1+X^2)^2} \quad \mbox{and} \quad \pp{\tilde{U}}{Z}\bigg|_{Z=0} = - \frac{6 X}{(1+X^2)^2}\,.
\end{equation}

As in \eqref{vdeltaexpansions}, we write $\bm{V} = \bm{V}_0 + \Lambda \bm{V}_1 + \Lambda^2 \bm{V}_2 + \O(\Lambda^3) $ and $\Delta(X) = -\Lambda P =  -\Lambda P_0 + \Lambda^2 P_1 + \Lambda^3 P_2 + \O(\Lambda^4)$, noting that we now utilize the solution for the flow up to $\O(\Lambda^2)$. Constraining the cylinder to translate parallel to the wall ($w_c = 0$), and substituting the expansions for $\bm{v}$ and $\Lambda$ into \eqref{ForceRecip}, we obtain
\begin{equation} \label{eq:angmom}
F^H_{\mathrm{lift}} \equiv \bm{F}^H \cdot \ez = F_0 + \Lambda F_1 + \Lambda^2 F_2 + \Lambda^3 F_3 +  \O(\Lambda^4)\,, 
\end{equation}
where 
\begin{subequations}
	\begin{align}
	F_0 &= - \frac{2 \mu u_c a}{h_f} \intinf \pp{\tilde{U}}{Z}\bigg|_{Z=0} \rmd X\,,\\
	F_1 &= - \frac{2 \mu u_c a}{h_f} \intinf \left\{\tilde{P} \pp{P_0}{X} - P_0 \pp{\tilde{U}}{Z} \pp{U_0}{Z}\right\}\Bigg|_{Z=0} \rmd X\,, \\
	F_2 &= - \frac{2 \mu u_c a}{h_f}\intinf\left\{\tilde{P} \pp{P_1}{X} - P_0 \left(\pp{\tilde{U}}{Z} \pp{U_1}{Z} + \tilde{P} \pp{U_1}{Z} \right) - P_1 \pp{\tilde{U}}{Z} \pp{U_0}{Z} + \frac{P_0^2}{2} \left( \pp{\tilde{U}}{Z} \pp{P_0}{X} + \tilde{P} \frac{\partial^2 U_0}{\partial X \partial Z}\right)\right\}\Bigg|_{Z=0} \rmd X,\\
	F_3 &= -\frac{2 \mu u_c a}{h_f} \intinf \Bigg\{ \tilde{P} \pp{P_2}{X} - P_0\left(\pp{\tilde{U}}{Z}\pp{U_2}{Z} + \tilde{P} \pp{U_2}{X} \right) - P_1\left(\pp{\tilde{U}}{Z}\pp{U_1}{Z} + \tilde{P} \pp{U_1}{X} \right) - P_2\pp{\tilde{U}}{Z}\pp{U_0}{Z} \nonumber \\
&\qquad\qquad + \frac{P_0^2}{2} \left(\pp{\tilde{U}}{Z} \pp{P_1}{X}  + \tilde{P} \frac{\partial^2 U_1}{\partial X \partial Z} \right) + P_0 P_1 \left(\pp{\tilde{U}}{Z} \pp{P_0}{X}  + \tilde{P} \frac{\partial^2 U_0}{\partial X \partial Z} \right) - \frac{P_0^3}{6} \tilde{P} \pp[2]{P_0}{X} \Bigg\}\Bigg|_{Z=0}\rmd X\,.
	\end{align}
\end{subequations}

Since $\partial{\tilde{U}}/\partial {Z}$ is an odd function of $X$, $F_0 = 0$ identically. At $\O(\Lambda)$, we recover the leading-order lift force result \citep{sek93,sko04}
\begin{equation}
F_1 = \frac{3\pi}{8} \frac{2 \mu u_c a}{h_f}\,.
\end{equation}
Again, by the odd symmetry of its integrand, we find $F_2 = 0$. Finally, the integral expression for $F_3$ evaluates to
\begin{equation}
	F_3 = - \frac{2 \mu u_c a}{h_f} \frac{3\pi(761 - 512 \beta_2)}{2048}.
\end{equation}
\begin{comment} 
\begin{align}
F_3 &= -\frac{\mu u_c \ell^2}{h_f^2} \intinf \bigg[ \tilde{P} \pp{P_2}{X} - P_0\left(\tilde{P} \pp{U_2}{X} + \pp{\tilde{U}}{Z}\pp{U_2}{Z}\right) - P_1\left(\tilde{P} \pp{U_1}{X} + \pp{\tilde{U}}{Z}\pp{U_1}{Z}\right) - P_2\left( \pp{\tilde{U}}{Z}\pp{U_0}{Z}\right) \nonumber \\
&\qquad\qquad + \frac{P_0^2}{2} \left(\pp{\tilde{U}}{Z} \pp{P_1}{X}  + \tilde{P} \frac{\partial^2 U_1}{\partial X \partial Z} \right) + P_0 P_1 \left(\pp{\tilde{U}}{Z} \pp{P_0}{X}  + \tilde{P} \frac{\partial^2 U_0}{\partial X \partial Z} \right) - \frac{P_0^3}{6} \tilde{P} \pp[2]{P_0}{X} \bigg]_{Z=0}\rmd X\,.
\end{align} 
where we have used the continuity and momentum equations to simplify some terms. Rearranging, we obtain the result
\begin{align}
F_3 & = \frac{\mu u_c \ell^2}{h_f^2} \intinf \bigg[\tilde{P} \left\{\pp{P_2}{X} - \left(P_0 \pp{U_2}{X} + P_1 \pp{U_1}{X} \right) + \left(\frac{P_0^2}{2} \frac{\partial^2 U_1}{\partial X \partial Z} + P_0 P_1 \frac{\partial^2 U_0}{\partial X \partial Z}\right) - \frac{P_0^3}{6}\pp[2]{P_0}{X} \right\} \nonumber \\
&\qquad + \pp{\tilde{U}}{Z} \left\{-\left(P_0 \pp{U_2}{Z} + P_1 \pp{U_1}{Z} + P_2 \pp{U_0}{Z}\right) + \left(\frac{P_0^2}{2} \pp{P_1}{X} + P_0 P_1 \pp{P_0}{X}\right)  \right\}\bigg]_{Z=0} \rmd X \nonumber \\
&= - \frac{\mu u_c \ell^2}{h_f^2} \frac{3\pi(761 - 512 \beta_2)}{2048}.
\end{align}
\end{comment}
The hydrodynamic lift force per unit length is therefore 
\begin{equation} \label{FliftApp}
F^H_{\mathrm{lift}} = \frac{2 \mu u_c a}{h_f} \left(\frac{3 \pi}{8} \Lambda - \frac{3\pi(761 - 512 \beta_2)}{2048} \Lambda^3 + \O(\Lambda^5)\right)\,.
\end{equation}
The above expression for the lift force is plotted in figure \ref{LiftForceFig}(b) for the torque-free case $(\beta_2 = \frac{21}{256})$. Note that the $\O(\Lambda^3)$ correction to the lift force is only weakly influenced by whether the cylinder is torque-free or is held under an external torque so that it does not rotate $(\beta_2 = 0 )$. Symmetry arguments also preclude additional drag or torque contributions at $\O(\Lambda^3)$.

\end{document}